\documentclass[10pt]{article}

\usepackage[T1]{fontenc}
\usepackage[utf8]{inputenc}
\usepackage[strict,autostyle]{csquotes}
\usepackage[USenglish]{babel}
\usepackage{microtype}
\usepackage{booktabs}
\usepackage{graphicx}
\usepackage{natbib}
\usepackage{authblk}
\usepackage{hyperref}

\graphicspath{{./figures/}}

\hypersetup{
	pdfauthor={Geoff Boeing},
	pdftitle={The Effects of Inequality, Density, and Heterogeneous Residential Preferences on Urban Displacement and Metropolitan Structure: An Agent-Based Model},
	pdfsubject={The Effects of Inequality, Density, and Heterogeneous Residential Preferences on Urban Displacement and Metropolitan Structure: An Agent-Based Model},
	pdfkeywords={urban science, agent-based model, gentrification, displacement, sprawl, land use, housing},
	pdffitwindow=true,         
	breaklinks=true,           
	colorlinks=false,          
	pdfborder={0 0 0}          
}

\begin{document}
	
\title{The Effects of Inequality, Density, and Heterogeneous Residential Preferences on Urban Displacement and Metropolitan Structure: An Agent-Based Model\footnote{Citation info: Boeing, G. 2018. \enquote{The Effects of Inequality, Density, and Heterogeneous Residential Preferences on Urban Displacement and Metropolitan Structure: An Agent-Based Model.} \textit{Urban Science}, 2 (3), 76. doi:10.3390/urbansci2030076}}
\date{August 2018}
\author[]{Geoff Boeing\thanks{Email: \href{mailto:g.boeing@northeastern.edu}{g.boeing@northeastern.edu}}}
\affil[]{School of Public Policy and Urban Affairs\\Northeastern University}

\maketitle

\begin{abstract}
Urban displacement---when a household is forced to relocate due to conditions affecting its home or surroundings---often results from rising housing costs, particularly in wealthy, prosperous cities. However, its dynamics are complex and often difficult to understand. This paper presents an agent-based model of urban settlement, agglomeration, displacement, and sprawl. New settlements form around a spatial amenity that draws initial, poor settlers to subsist on the resource. As the settlement grows, subsequent settlers of varying income, skills, and interests are heterogeneously drawn to either the original amenity or to the emerging human agglomeration. As this agglomeration grows and densifies, land values increase, and the initial poor settlers are displaced from the spatial amenity on which they relied. Through path dependence, high-income residents remain clustered around this original amenity for which they have no direct use or interest. This toy model explores these dynamics, demonstrating a simplified mechanism of how urban displacement and gentrification can be sensitive to income inequality, density, and varied preferences for different types of amenities.
\end{abstract}

\section{Introduction}

Prosperous, affluent cities have become prominent battlegrounds in the public debate around gentrification and displacement of the poor \citep{zuk_gentrification_2018,newman_right_2006}. Cities and neighborhoods often initially develop around a spatial amenity, such as the harbors of San Francisco and New York, that provides subsistence to initial residents. Subsequent newcomers might be attracted to this growing urban agglomeration, bringing to the city new skills, education, wealth, and higher land values that price out the poor, pushing them to the less accessible periphery. Patterns of wealth concentration, suburbanization, and displacement are well-known to urban scholars \citep{torrens_modeling_2007,zuk_gentrification_2018,smith_toward_1979,newman_right_2006}. However, the underlying dynamics of path dependence and heterogeneous residential preferences of different classes remain under-explored.

This paper presents an exploratory agent-based model (ABM) of urban settlement, agglomeration, and displacement. It demonstrates a simplified conceptual mechanism by which the rich might displace the poor away from some spatial amenity on which the latter rely, despite the former having no direct interest in this amenity. In this model, settlements form around spatial amenities that draw initial settlers who subsist on this resource. As the settlement grows, subsequent settlers of varying income, skills, and interests will be attracted to either (1) the original amenity or (2) the emerging human agglomeration. As the agglomeration grows and densifies, land prices increase, and the initial poor settlers are displaced from the amenity on which they rely. 

This study uses this toy model's simplified mechanism to illustrate and consider urban policy scenarios. How are settlement patterns affected if different classes of people have differing preferences correlated with their ability to pay? How might income inequality and density affect diversity and displacement as land values rise? How can path dependence impact city spatial structure and the poor's access to the resources on which they rely? The ABM presented here is a toy model to conceptually elucidate how such processes might evolve. This paper first discusses the theoretical framework of spatial amenities, urban agglomeration, displacement, and path dependence. Then it presents the study's research questions, model overview, model components, parameters, and state variables. Next it describes the simulation results, including a sensitivity analysis and discussion of findings, before concluding.

\section{Theoretical Framework}

\subsection{Spatial Amenities}

Spatial amenities draw people to a place \citep{mumford_city_1961} and can include both natural resources and social amenities. Prior to city formation and urban agglomeration, early residents in an area might settle near natural amenities that provide some relative spatial advantage \citep{arthur_urban_1988}---for example, sea coasts for transportation purposes, rivers for water and energy, or hilltops for defense \citep{kostof_city_1991,kostof_city_1992}. Potential farmers might be drawn to fertile, well-irrigated land for agricultural purposes. Other natural resources might be useful for extraction and transformation, such as ore and timber.

While initial settlers may be drawn to certain places in unpopulated landscapes because of their high natural amenity value, subsequent settlers might be drawn to that same point for two reasons \citep{mumford_city_1961}. First, they may also be interested in the same natural amenities that drew the first wave of settlers. Or second, they may instead be interested in the nascent human agglomeration that has formed around these natural amenities. People drawn to the urban amenity (human concentration) might have no direct interest whatsoever in this specific natural amenity---it is simply an irrelevant spatial covariate of initial human presence and urban agglomeration.

Today, examples of urban spatial amenities include land, housing, and access to markets and services \citep{osullivan_urban_2008}. This paper abstractly models natural spatial amenities that are club goods, such as ideal locations in a harbor for a port. Such resources are non-extractive and non-rivalrous in that human use does not diminish their stocks. However they are at least indirectly excludable in that private property land markets can arise on and around them. Thus, access is metered by willingness and ability to pay. For the purposes of this study, abstract natural amenities are modeled, with preferences to live near them correlated with income.

\subsection{Urban Agglomeration}

Following the early work of \citet{thunen_isolierte_1826}, \citet{christaller_central_1933}, and \citet{alonso_location_1964} on city structure, a large body of urban scholarship today argues that cities exist because people want to be near one another for social, economic, or political reasons \citep{harris_nature_1945,jacobs_economy_1969,phe_status_2000,glaeser_triumph_2011,moretti_new_2013}. Urban agglomerations form when people buy and sell goods and services while minimizing travel costs and distances \citep{boarnet_travel_2001}. They live near one another to take advantage of shared infrastructure and cultural amenities: firms and workers benefit from knowledge spillovers, labor pooling and matching, and information sharing. These forces of agglomeration cause the centralization that leads to urbanization and the ongoing spatial distribution of location choices and infrastructure investment within cities \citep{chatman_public_2011}.

\citet{osullivan_urban_2008} explains how households and firms individually choose where to locate as a function of destination accessibility because transportation to amenities requires time, money, and resources. Each location choice depends on an individual evaluation and trade-off between transportation costs and spatial demands. Accordingly, land and housing costs tend to decline with distance from wherever people want to be in great numbers, and in turn people choose residential locations by trading off accessibility and housing costs \citep{rodriguez_can_2014}. Density is thus greatest in highly accessible urban centers where people trade off housing size for access \citep{osullivan_urban_2008}. However, such choices can be restricted in modern cities with low-density zoning schemes that limit people's abilities to economize on space \citep{lens_strict_2016}.

Urban economists and geographers have studied urban agglomeration in terms of economies and dis-economies of scale \citep{krugman_increasing_1991,krugman_new_2011}. Economies of scale---such as low transportation costs, shared infrastructure, labor pooling, job matching, information sharing, and knowledge spillovers---produce centripetal forces that stimulate centralization and agglomeration \citep{fujita_economics_1996}. Dis-economies of scale---such as crime, traffic congestion, and high rents and costs of living---produce centrifugal forces that promote decentralization away from cities' cores. According to this theory, these competing push-pull forces over time explain the eventual spatial distribution of a metropolitan area \citep{krugman_self_1996,duranton_nursery_2001}.

\subsection{Displacement and Path Dependence}

Urban displacement occurs when a household is forced to relocate due to conditions affecting its home or surroundings, and is often caused by rising housing costs in prosperous cities \citep{zuk_gentrification_2018,smith_toward_1979,smith_new_2002,lees_gentrification_2016,grodach_gentrification_2018}. Displacement and gentrification can result from planning and economic development decisions, and play a significant role in debates around livability, social justice, and equity \citep{zuk_housing_2016,newman_right_2006,hwang_divergent_2014}. \citet{torrens_modeling_2007} explain that gentrification and displacement---much studied in urban geography, economics, and sociology for the past 50 years---encompass the \enquote{transition of property markets from relatively low value platforms to higher value platforms under the influence of redevelopment and influx of higher-income residents, often with spatial displacement of original residents and an associated shift in the demographic, social, and cultural fabric of neighborhoods under its influence.}

An important factor in this phenomenon is path dependence, when a system's state depends heavily on its previous state \citep{arthur_urban_1988}. For example, positive feedback loops, sunk costs, clustering, and agglomeration can lead to inefficient spatial outcomes and circumscribed location choices, even if the original circumstances that determined earlier behavior are no longer relevant \citep{geest_history_2000}. In cities, historical accidents and decisions from long ago can shape the trajectory of urban systems deep into the future \citep{allen_urban_1981}. Certain system characteristics may eventually exist merely as vestigial artifacts of this past, with no other rationale for their present existence. In cities, spatial characteristics may have high fixed costs, making it prohibitively expensive to pick up and start over again.

For example, consider the first settlers to a region, selecting homesteads near the spatial amenities they require to support their livelihoods and survival cf. \citep{hoyt_homer_1954}. This might be a natural harbor ideal for shipping and transportation. Certain subsequent settlers might also locate nearby in pursuit of the same amenities offered by this harbor. However, others might now settle near this harbor because they want to exploit the labor pool or provide second-order value-added services to the local residents. These newcomers might be bankers, teachers, doctors, or administrators. Their high-skill services require education and experience, and they may have considerably higher incomes than the first wave of residents. These newcomers only settled at this spatial location because they wanted to be near the center of human agglomeration to provide their services efficiently. They have no direct interest in the harbor that happens to be there. Nevertheless, as they compete for limited space, their income and ability to pay increases land prices and rents \citep{osullivan_urban_2008}.

In this scenario, rising costs displace poor longtime residents from central neighborhoods by pushing land prices at the center of the agglomeration beyond their abilities to pay cf. \citep{smith_toward_1979,hoyt_homer_1954}. The poor may relocate to affordable land at the periphery, far from the amenity that catalyzed their initial settlement. Over time, the high-income residents remain clustered at the center around the natural amenity while the poor are pushed ever farther from the spatial resources they rely on. To conclude this illustrative example, the eventual spatial distribution of this urban area will not be \emph{efficient} defined as when no one can be made better off without making someone else worse off \citep{newbery_pareto_1984,mathur_how_1991,hatna_combining_2015}. In our scenario, the rich have no direct use for the specific natural amenity they remained clustered around---they only remain there due to path dependence. They have no incentive to move away from this point of high urban amenity, so they simply remain in place. The poor would be better off were they closer to these natural amenities, but they can no longer afford to live there.

Location decisions result from heterogeneous preferences being acted on in complex, nonequilibrium urban systems \citep{batty_new_2013,johnson_cities_2017,boeing_nonlinear_2016}. Simulation models usefully represent the endlessly evolving outcomes of such collective dynamics \citep{white_cellular_1993,heppenstall_space_2016}. \citet{brown_effects_2006} developed an ABM with heterogeneous residential location preferences and found that agents preferring aesthetics sprawled more than agents preferring proximity. \citet{hatna_combining_2015} used an ABM to study Schelling segregation dynamics when agents have heterogeneous within-group preferences. \citet{axtell_agent-based_1994} explain that ABMs demonstrate how social structures collectively grow out of interactions between individuals. \citet{parker_multi-agent_2003} suggest that multi-agent spatial models are \enquote{particularly well suited for representing complex spatial interactions under heterogeneous conditions and for modeling decentralized, autonomous decision making.} \citet{torrens_modeling_2007} build an agent-based cellular model of gentrification to explore the displacement of original residents and changes in population composition, arguing that these methods are well-suited to illustrating these processes.

\section{Methods}

Accordingly, in this study, we implement a spatial agent-based toy model to consider a conceptual mechanism by which the rich might displace the poor from a spatial amenity on which the latter rely, despite the former having no direct interest in this amenity. This simulates the outcome of many heterogeneous individuals interacting under bounded rationality. \citet{axtell_agent-based_1994} classify ABMs into four levels of analysis. Within their classification system, this toy model is designed at level 0, with some characteristics of level 1---that is, it provides a simplified conceptual representation of reality to demonstrate a theoretical mechanism, but qualitatively shows concordance with real-world expectations.

This study poses three questions. First, how are settlement patterns affected if different classes of people have differing preferences correlated with their ability to pay (e.g., the poor prefer natural amenities but the rich prefer urban amenities)? We suspect that the higher-income agents will displace the poor agents from the accessible center to the inexpensive periphery. Second, how do income inequality and density affect diversity and displacement of the poor as housing costs rise? We suspect that as the rich, middle, and poor incomes are increasingly stratified, the poor are more likely to be displaced and diversity decreases. In other words, as income equality increases, diversity increases as agents of different classes can more easily live together without the middle-income or poor being priced out. Third, how does path dependence affect the spatial distribution of the urban area and the poor's access to the physical resources on which they subsist? We suspect that after the poor are displaced, the rich agents remain clustered around these natural amenities---even though they have no direct preference for them---due to path dependence.

\subsection{Model Overview and Purpose}

To probe these questions, this study built an ABM with NetLogo 5.1. It serves as a simplified conceptual model that demonstrates how residential displacement can occur given multiple parameters. Here we borrow from Grimm et al.'s \citep{grimm_odd_2010,grimm_standard_2006} \enquote{Overview, Design concepts, and Details} (ODD) model description protocol to describe this model. At a high level, the basic operation of the model comprises four simple steps:

\begin{enumerate}
	\item Abstract natural amenities are distributed across a landscape. This amenity is a club good.
	\item Poor agents flock to these amenities and in turn create small initial human/urban agglomerations.
	\item Higher-income agents flock to these incipient urban agglomerations to take advantage of the growing urban amenity, provide high-skill value-added services, and exploit available labor.
	\item Land values increase as a function of the count and income of agents present on a cell. If a cell exceeds an agent's budget, the agent relocates to the most desirable affordable cell.
\end{enumerate}

\subsection{Model Components}

This model consists of two types of components: agents and cells. Agents represent households and come in three subsets representing socioeconomic classes: rich, middle-income, and poor. The rich agents prefer urban amenities and are colored yellow in the model's world. The middle-income agents also prefer urban amenities but have smaller incomes than the rich agents. They are colored orange in the model's world. Lastly, the poor agents prefer natural amenities and have yet lower incomes. They are colored red.

The cells compose a $33 \times 33$ grid. Each cell represents an acre (thus the grid approximates the size of a town) and has two types of possible amenity value. The first is the natural amenity value, which is distributed across the landscape at setup according to the parameters described in the subsequent section. This study emphasizes natural amenities concentrated at a single point. The second is the urban amenity value, which represents the urban agglomeration on or near the cell as a function of the number of agents present and their income. Thus, rich agents contribute more to urban agglomeration than middle-income agents or poor agents do: rich agents have larger incomes to invest in the creation and sustenance of urban, economic, and cultural institutions. Each cell also has a cost to live there (as a function of its urban amenity value), and any agent interested in settling must have an income that exceeds this cost. This reflects the capitalization of urban agglomeration amenity value (through accessibility) and agent willingness to pay. As land uses and values do not mutate themselves \citep{bandini_cellular_2001}, this is a cellular model in which agents' actions result in changes to socially-constructed traits of the underlying land.

\subsection{Model Parameters and Constants}

The model is controlled with 13 parameters (Table \ref{tab:parameters}) chosen from the urban studies literature on housing choice and displacement, focusing on housing rather than infrastructure such as water or transportation (cf. theoretical framework section and \citep{chatman_residential_2009,desmond_eviction_2012,chatman_role_2013,zuk_gentrification_2018}). The \textit{initial number of agents} parameter specifies the number of agents present upon initializing the model. All of these agents are poor and they are randomly distributed across the landscape. After these initial agents settle, a new agent is spawned each time step until the carrying capacity is met. The \textit{regional carrying capacity} specifies the maximum possible population of agents in the world. Once this limit is reached, no new agents are spawned. The \textit{maximum density} parameter specifies the maximum number of agents that may occupy a single cell at any time cf. \citep{southworth_walkable_1997}.

\begin{table}[h]
	\centering
	\caption{Model parameters and summaries. See accompanying text for a complete description of each.}
	\label{tab:parameters}
	\scriptsize
	\begin{tabular}{lll}
		\toprule
		\textbf{Parameter }                 & \textbf{Brief Description}                                     & \textbf{Range}      \\
		\midrule
		Initial number of agents   & Number of poor to randomly distribute at setup        & 0--100    \\
		Regional carrying capacity & Max population of agents in the model                 & 1--2000   \\
		Maximum density            & Max possible agents on a single cell at a time        & 1--100    \\
		Poor income                & Basis for calculating incomes of poor agents          & 1--100    \\
		Middle income              & Basis for calculating incomes of middle-income agents & 1--200    \\
		Rich income                & Basis for calculating incomes of rich agents          & 1---200    \\
		Seek radius                & Search radius when agent is in seek mode              & 1--100    \\
		Relocate radius            & Search radius when agent is in settled mode           & 1--20     \\
		Diffusion distance         & Number of iterations of amenity value diffusion       & 0--20     \\
		Diffusion rate             & Amount of amenity value to diffuse in each iteration  & 0--1      \\
		Single amenity             & Distribute natural amenities randomly, or not         & True/False \\
		Spawn middle               & Spawn middle-income agents, or not                    & True/False \\
		Spawn poor                 & Spawn poor agents after initial setup, or not         & True/False \\
		\bottomrule
	\end{tabular}
\end{table}

The \textit{income inequality} meta-parameter combines three parameters that define the agents' incomes. Each poor agent's income is randomly sampled from a distribution between zero and the poor income parameter value. Each middle-income agent's income is randomly sampled from a distribution between the poor income and the middle income parameter values. Each rich agent's income is randomly sampled from a distribution between the middle income and the rich income parameter values. After spawning, agents seek the grid cell with the highest value of their preferred amenity and a cost below their income. An agent's \textit{search radius} is defined by two parameters corresponding to different types of searches. The \textit{seek radius} defines the set of cells an agent considers moving to when it is in seek mode (i.e., it has not settled down). The \textit{relocate radius} defines the set of cells an agent considers moving to when it is currently settled down and is considering relocating because housing costs exceed its budget. An agent is considered \enquote{settled down} if it has not moved for at least 10 time steps.

Accessibility is modeled by smoothing the value of natural amenities and urban agglomeration across the landscape by way of two parameters. The \textit{diffusion distance} parameter specifies how many iterations of diffusion occur and the \textit{diffusion rate} parameter specifies what proportion of the cell's value is diffused to neighboring cells during each of these iterations. The \textit{single amenity} parameter is a Boolean that defines how natural amenities are distributed across the landscape at initial setup. When true, natural amenities are distributed randomly before diffusion. When false, natural amenities are consolidated at the center of the landscape, with only trivial amounts of natural amenity distributed randomly elsewhere.

The \textit{spawn middle} parameter defines whether or not middle-income agents will be spawned during the simulation. If false, the world will only have rich and poor agents. The \textit{spawn poor} parameter defines whether or not poor agents will be spawned during the simulation. If both of these spawning parameters are false, only rich agents will be spawned after the initialization of the model.

\subsection{State Variables}

The model has three state variables which define the overall state of the system at any given time (Table \ref{tab:state_variables}). The \textit{number of agents} variable straightforwardly indicates the count of agents currently in existence in the model world. It iterates whenever a new agent spawns. The \textit{diversity indicator} measures how diverse the agents' settlement patterns are. To calculate it, we first calculate (for each agent) the proportion of agents within a Chebyshev distance of 1 that are of a different class (rich, middle-income, poor) than this agent, and then determine the mean of these proportions. This provides an indicator of diversity---that is, the proportion of \enquote{other} kinds of agents that the average agent is exposed to within its Moore neighborhood. For example, a diversity indicator of 0.5 means that, on average, half of an agent's neighbors are of a different class than the agent.

\begin{table}[h]
	\centering
	\caption{State variables and summaries. See accompanying text for a complete description of each.}
	\label{tab:state_variables}
	\begin{tabular}{ll}
		\toprule
		\textbf{State Variable}        & \textbf{Brief Description}                                  \\
		\midrule
		Num. of agents        & Current number of agents in the world              \\
		Diversity indicator   & Diversity of the agents' settlement patterns    \\
		Poor access indicator & Quality of the poor's access to desired amenities \\
		\bottomrule
	\end{tabular}
\end{table}

The \textit{poor access indicator} measures the accessibility of the poor agents to natural amenities. To calculate it, we first calculate (for each poor agent) the ratio of its current cell's natural amenity value to the maximum natural amenity value of any cell that was ever within this agent's seek radius. Then we calculate the mean of these ratios. This provides an indicator of how \enquote{good} the average poor agent's access is to natural amenities and serves as a proxy for efficiency. For example, an indicator value of 1.0 means that all poor agents are currently on the cell (within their seek radii) with the highest natural amenity value. Thus, each agent is on its most preferred cell. An indicator value of 0.5 means that the average poor agent is on a cell with only half the natural amenity value of the best cell that this agent has seen and would prefer to be on, if it could afford it.

\subsection{Parameterization and Simulation}

In this study we hold six of these parameters constant (Table \ref{tab:constants}). The initial number of agents is 100, the regional carrying capacity is 1000, spawn middle is true, and spawn poor is false. Thus the model will always initialize with 100 poor agents (the initial settlers) just to simulate the initial regional population pre-urbanization. Once they have settled, a rich or middle-income agent (randomly one or the other) will spawn once per time step until the population reaches 1000. Once the last agents settle, the simulation ends. This study demonstrates a mechanism by which an existing set of low income residents might be displaced away from some amenity through class succession (it does not attempt to model the arrival of new low-income immigrants), thus it sets the spawn poor parameter to false. Finally, accessibility is modeled by setting the diffusion distance to 10 and the diffusion rate to 0.5.

\begin{table}[h]
	\centering
	\caption{Model constants and their values.}
	\label{tab:constants}
	\begin{tabular}{ll}
		\toprule
		\textbf{Constant}                   & \textbf{Value} \\
		\midrule
		Initial number of agents   & 100   \\
		Regional carrying capacity & 1000  \\
		Diffusion distance         & 10    \\
		Diffusion rate             & 0.5   \\
		Spawn middle               & True  \\
		Spawn poor                 & False \\
		\bottomrule
	\end{tabular}
\end{table}

We first explore the model by simulating it with four sets of parameter values (Table \ref{tab:parameterizations}). For each set, we run the simulation 50 times, initialized with randomly seeded agents and amenities, and then calculate the means and confidence intervals of the state variables. 50 runs is the point at which convergence can be safely and consistently assumed as the coefficient of variation remains nearly the same after each subsequent run cf. \citep{lorscheid_opening_2012,thiele_facilitating_2014}. For the purposes of this study, we use these four illustrative scenarios based on standard scenario planning practice. These scenarios are parameterized according to the empirical literature in urban planning, sociology, and demography e.g., \citep{southworth_walkable_1997,u.s._census_bureau_income_2011,desmond_eviction_2012}.

\begin{table}[h]
	\centering
	\caption{Parameter values for each set of exploratory simulations. Set A is the baseline with moderate values, Set B represents high income inequality, Set C represents low density, and Set D represents dispersed natural amenities and wide relocation search radii. The values of the parameters held constant appear in Table \ref{tab:constants}.}
	\label{tab:parameterizations}
	\begin{tabular}{lllll}
		\toprule
		\textbf{Parameter }      & \textbf{Set A} & \textbf{Set B} & \textbf{Set C} & \textbf{Set D} \\
		\midrule
		Maximum density & 25    & 25    & 4     & 25    \\
		Poor income     & 20    & 5     & 20    & 20    \\
		Middle income   & 50    & 100   & 50    & 50    \\
		Rich income     & 100   & 200   & 100   & 100   \\
		Single amenity  & True  & True  & True  & False \\
		Relocate radius & 3     & 3     & 3     & 20    \\
		\bottomrule
	\end{tabular}
\end{table}

Simulation Set A---the baseline---represents moderate land use restrictions (i.e., a density constraint) and income inequality, and centralizes the natural amenity in the landscape. The parameter values are drawn from existing empirics and literature cf. \citep{schmolke_ecological_2010}. The maximum density of 25 dwelling units per acre represents medium density development, such as apartments or rowhouses cf. \citep{southworth_walkable_1997}. The relocation search radius reflects a desire to not be displaced, seeking new housing only within one's general vicinity cf. \citep{desmond_eviction_2012}. Income inequality is parameterized with a poor income of 20, a middle income of 60, and a rich income of 100: these would approximate the 20th, 50th, and 80th percentile household incomes in thousands of US dollars in the 2010 US Census \citep{u.s._census_bureau_income_2011}. However, note that the agents' income distribution after running a simulation depends on the numbers of each type of agent spawned.

Set B uses the baseline values from Set A, but increases income inequality to reflect a highly unequal distribution of income. The poor, middle, and rich incomes would approximate the 3.5th, 80th, and 96th percentile household incomes, respectively, in the 2010 US Census. Set C uses the baseline parameter values from Set A, but decreases the maximum density to reflect a more restrictive exclusionary zoning regime by allowing only four dwelling units per acre---typical of a sprawling North American single-family suburb with large lot sizes \citep{southworth_walkable_1997}. Finally, Set D uses the baseline parameter values from Set A, but distributes natural amenities more evenly across the landscape and increases the agents' relocation search radius, allowing them to relocate to distant locations across the region. These four sets constitute an initial exploration of the model---we systematically explore higher/lower densities and greater/lesser income equality (key parameters of interest) in the subsequent sensitivity analysis.

\section{Results}

\subsection{Simulation Set A: Baseline Parameterization}

\begin{figure}[h]
	\centering
	\includegraphics[width=1.00\textwidth]{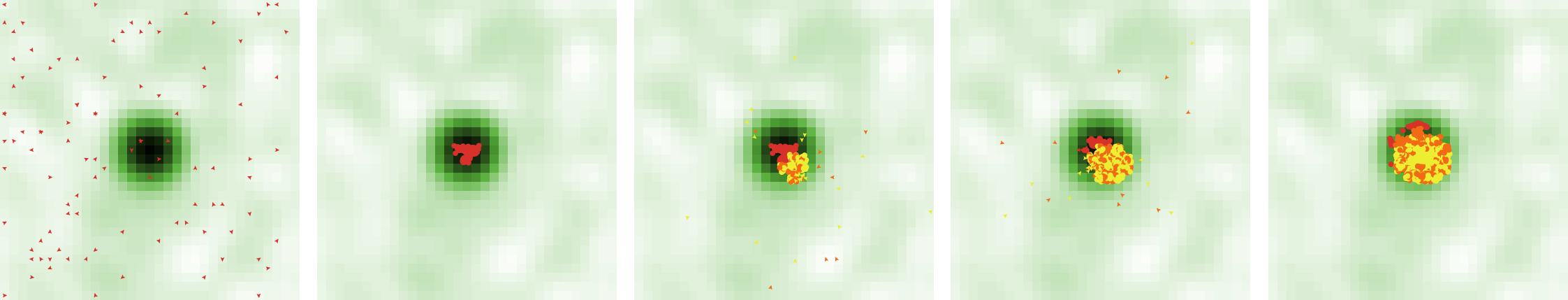}
	\caption{Evolution of Set A over time, from initialization (left) through end of simulation (right).}
	\label{fig:set_a_map}
\end{figure}

\begin{figure}[h]
	\centering
	\includegraphics[width=1.00\textwidth]{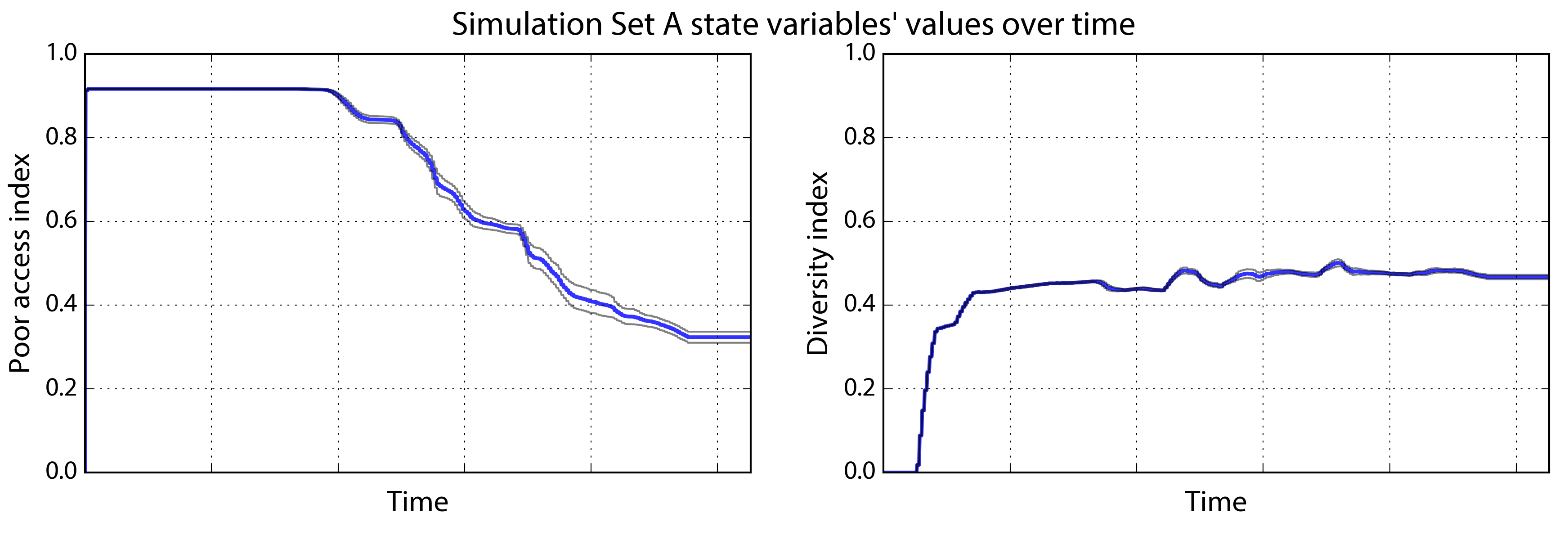}
	\caption{Set A's state variables from initialization through the end of the simulation, with mean value (blue) and 95\% confidence interval (gray) across 50 runs.}
	\label{fig:set_a_chart}
\end{figure}

We first test the model on a landscape with a centralized point of natural amenity and set the parameter values as defined in Table \ref{tab:parameterizations}. Upon setup, we see the centralized natural amenity in the landscape and 100 poor agents distributed randomly across it (Figure \ref{fig:set_a_map}). As the simulation runs, the poor agents converge on the center to be near the natural amenity. At this point, the diversity indicator is 0.00 and the poor access indicator is 0.91 (Figure \ref{fig:set_a_chart}). The settlement is completely homogenous and the poor agents have good access to the natural amenities they seek: the typical agent is on a patch with 91\% the natural amenity value of its preferred patch. The convergence of poor agents forms a small human agglomeration, increasing the urban amenity value of the central cells.

\unskip

As rich agents and middle-income agents arrive, they are drawn to these central cells for their urban amenity value, not their natural amenity value. At this point, the diversity indicator has risen to 0.40 and the poor access indicator remains 0.91. There is more diversity now as the rich have moved in: 40\% of a typical agent's neighbors are of a different class than the agent. Poor access still remains high as there are not enough rich or middle-income agents present to raise costs and displace the poor from their settlement. But as rich agents continue to flock to the urban agglomeration in increasing numbers, they begin to drastically raise the urban amenity value of these central cells. This makes the center both more attractive to subsequent rich and middle-income agents and increasingly unaffordable for poor agents.

The poor agents are soon forced to relocate to the periphery to find cells within their budget. Such cells are fairly undesirable for these agents---they are merely an affordable last option. At the end of the simulation, the diversity indicator has risen slightly to 0.47 and the poor access indicator has fallen to 0.32 (Table \ref{tab:convergence}), indicating that 47\% of the average agent's neighbors are of a different class than the agent and that the average poor agent is on a patch with just 32\% the natural amenity value of its preferred patch. The poor access indicator is relatively low as the poor agents have been displaced towards the amenity-deficient periphery of the settlement.

\begin{table}[htbp]
	\centering
	\caption{State variables' mean convergence values and 95\% confidence intervals across 50 runs for each set of exploratory simulations. Set A represents the baseline, Set B represents high income inequality, Set C represents low density, and Set D represents dispersed amenities and wide relocation radii.}
	\label{tab:convergence}
	\begin{tabular}{lrrrr}
		\toprule
		& \multicolumn{2}{c}{\textbf{Poor Access Indicator}}  & \multicolumn{2}{c}{\textbf{Diversity Indicator}}  \\
		\midrule
		& \textbf{Mean}                & \textbf{95\% Conf.} & \textbf{Mean}  & \textbf{95\% Conf.} \\
		\midrule
		Set A                 & 0.323               & 0.310--0.337   & 0.467 & 0.462--0.472   \\
		Set B                 & 0.124               & 0.117--0.132   & 0.425 & 0.420--0.431   \\
		Set C                 & 0.103               & 0.101--0.104   & 0.042 & 0.039--0.045   \\
		Set D                 & 0.865               & 0.840--0.891   & 0.433 & 0.428--0.439   \\
		\bottomrule
	\end{tabular}
\end{table}

\subsection{Simulation Set B: High Inequality}

\begin{figure}[h]
	\centering
	\includegraphics[width=1.00\textwidth]{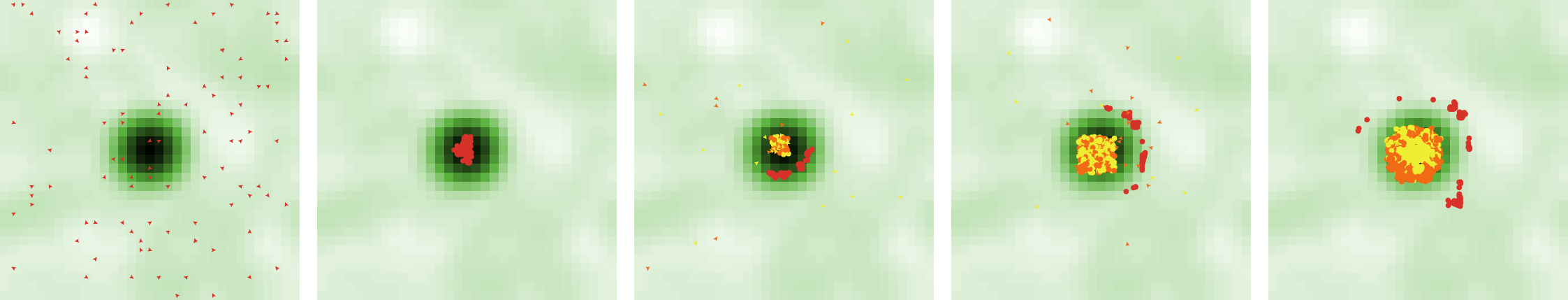}
	\caption{Evolution of Set B over time, from initialization (left) through end of simulation (right).}
	\label{fig:set_b_map}
\end{figure}

\begin{figure}[h]
	\centering
	\includegraphics[width=1.00\textwidth]{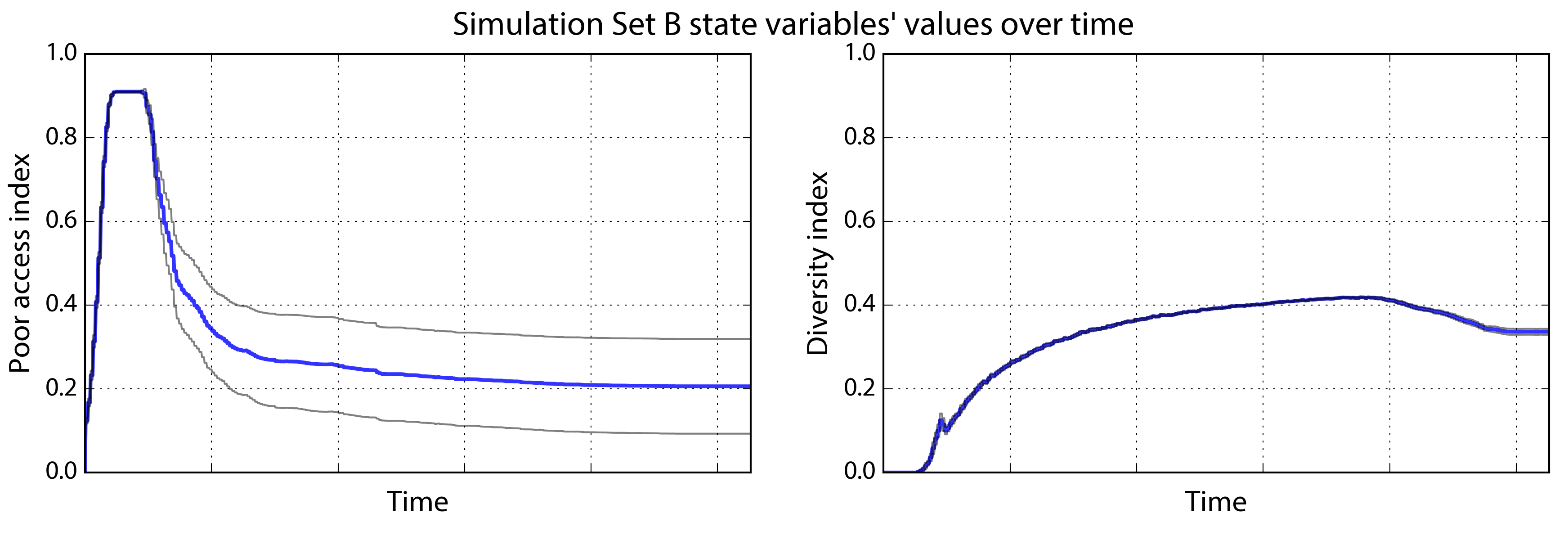}
	\caption{Set B's state variables from initialization through the end of the simulation, with mean value (blue) and 95\% confidence interval (gray) across 50 runs.}
	\label{fig:set_b_chart}
\end{figure}

Next, we increase income inequality by setting the parameter values as defined in Table \ref{tab:parameterizations} for Set B. Similar to the previous simulation, we initially see the centralized natural amenity in the landscape and 100 poor agents distributed randomly across it (Figure \ref{fig:set_b_map}). The poor agents converge on the center to be near the natural amenity, forming a small agglomeration and increasing the urban amenity value of the central cells. Again the diversity indicator is now 0.00 and the poor access indicator is 0.91 (Figure \ref{fig:set_b_chart}).

Rich and middle-income agents begin to settle near the small urban agglomeration of poor agents, displacing them much more quickly this time. The rich and middle incomes are far greater, creating higher land costs that the poor---with their far lower incomes---rapidly cannot afford. By the end of the simulation, the poor have been pushed further toward the periphery and away from their desired natural amenity. As in Set A, the dense yellow center of the urban core is predominantly rich agents, surrounded tightly by a suburban ring of mixed rich and middle-income agents. The poor access indicator and diversity indicator converge to 0.12 and 0.43 respectively.

\unskip

\subsection{Simulation Set C: Low-Density Zoning}

\begin{figure}[h]
	\centering
	\includegraphics[width=1.00\textwidth]{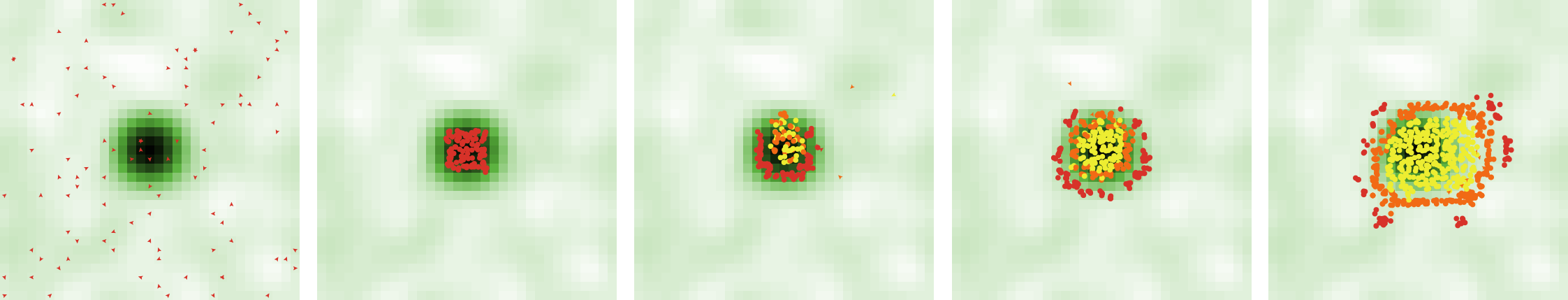}
	\caption{Evolution of Set C over time, from initialization (left) through end of simulation (right).}
	\label{fig:set_c_map}
\end{figure}

\begin{figure}[h]
	\centering
	\includegraphics[width=1.00\textwidth]{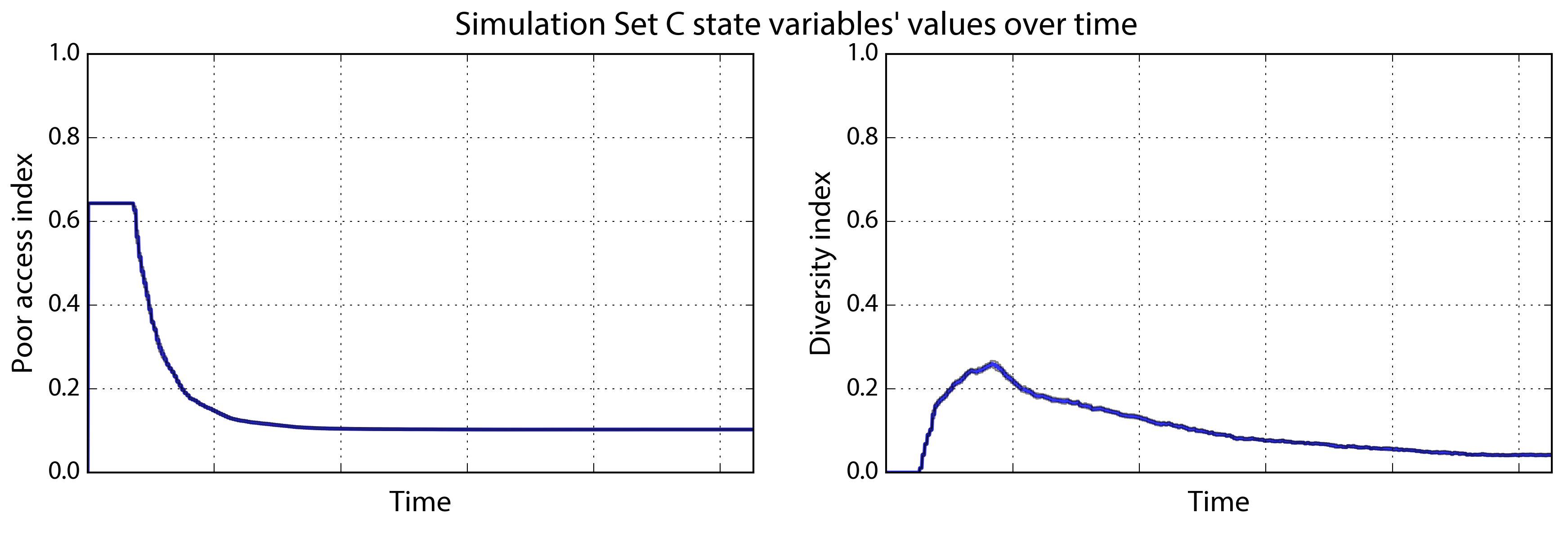}
	\caption{Set C's state variables from initialization through the end of the simulation, with mean value (blue) and 95\% confidence interval (gray) across 50 runs.}
	\label{fig:set_c_chart}
\end{figure}

Next, we return the income parameters to their baseline values but decrease the maximum density as defined in Table \ref{tab:parameterizations} for Set C. When the poor agents first converge on the natural amenity at the center, many of them remain dispersed around it due to the low density restrictions. This results in a relatively low initial poor access indicator of only 0.64. Further, the poor access indicator rapidly declines as the first wave of rich and middle-income agents immediately displaces the poor. Due to the density restrictions, the poor are pushed particularly far away from the natural amenity and few agents of any class can settle on their preferred cell (Figures \ref{fig:set_c_map} and \ref{fig:set_c_chart}).

\unskip

This simulation produces sprawling settlement patterns with no dense urban cores or centers of agglomeration. Diversity is low as the settlement quickly segregates according to agent ability to pay. At the center lies a loose mass of rich agents, surrounded by a thin suburban belt of middle-income agents.  Further out lie sprawling patches of poor agents, forced to the distant periphery where they could afford land.

\subsection{Simulation Set D: Dispersed Amenities}

For the next set of simulations, we set the density and income parameters to their baseline values but disperse natural amenities across the landscape and increase the relocation radius to 20. These parameter values are listed in Table \ref{tab:parameterizations}. When the poor agents converge on the center, their initial poor access indicator is very high---0.98---because even those agents who cannot reside on the centermost cell due to density restrictions still are on high amenity patches because of the dispersion of natural amenities across the landscape. The poor access indicator drops only slightly over time as rich and middle-income agents arrive, because the poor are displaced to other cells with decent natural amenities elsewhere.

\begin{figure}[h]
	\centering
	\includegraphics[width=1.00\textwidth]{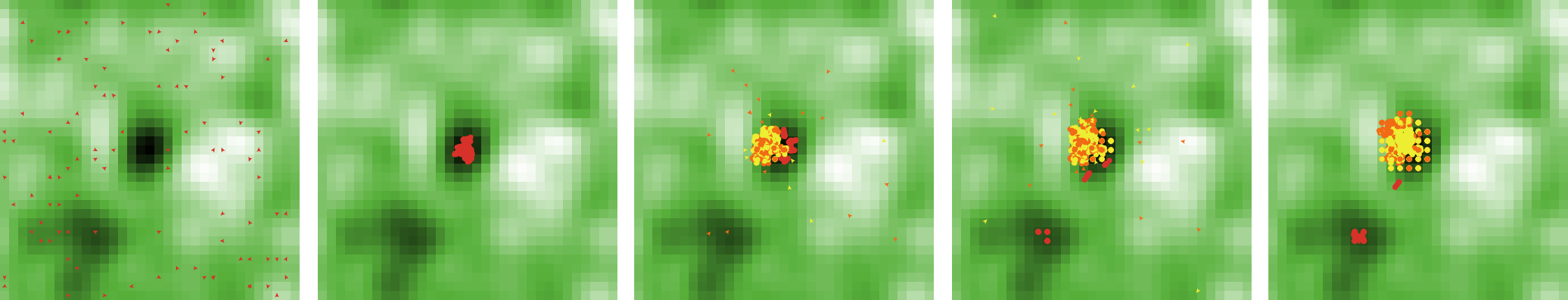}
	\caption{Evolution of Set D over time, from initialization (left) through end of simulation (right).}
	\label{fig:set_d_map}
\end{figure}

\begin{figure}[h]
	\centering
	\includegraphics[width=1.00\textwidth]{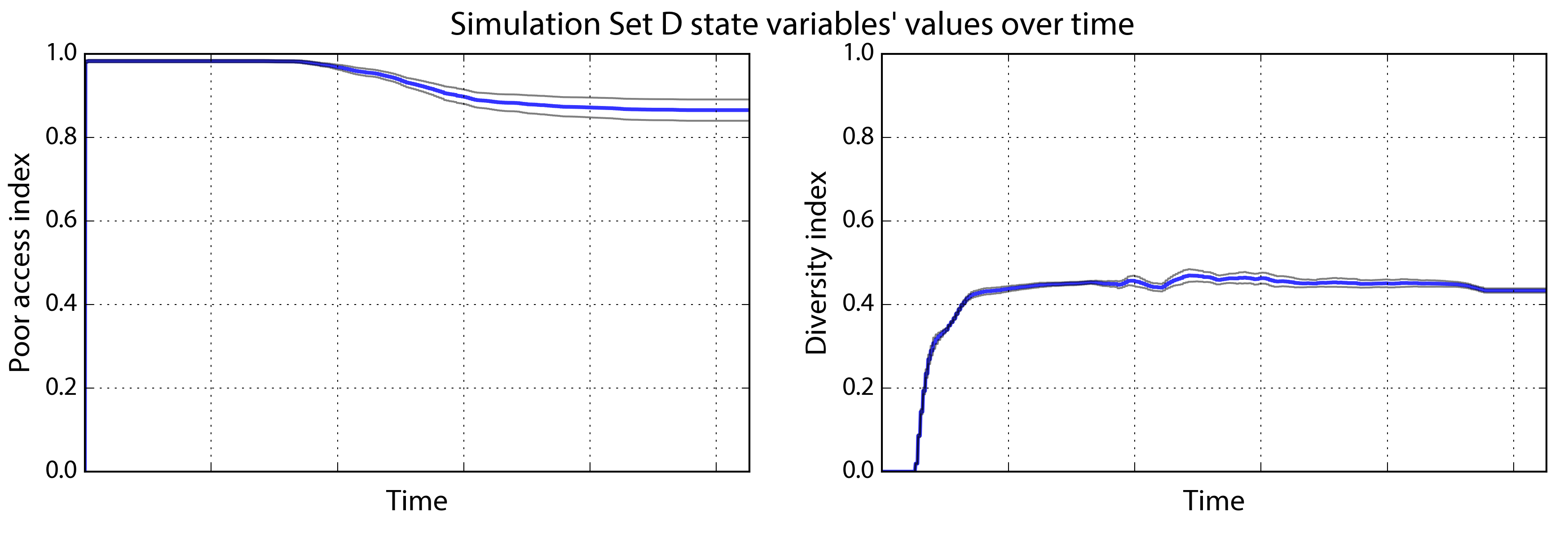}
	\caption{Set D's state variables from initialization through the end of the simulation, with mean value (blue) and 95\% confidence interval (gray) across 50 runs.}
	\label{fig:set_d_chart}
\end{figure}

In particular, as the central urban agglomeration grows in size and cost, the poor agents begin relocating to a new patch of high natural amenity to the southwest (Figure \ref{fig:set_d_map}). Although it is a second-best option, this patch allows the poor agents to maintain a much higher access indicator than in the previous simulation sets, due to the dispersion of amenities and the relocation radius. The poor access indicator converges at 0.87 (Figure \ref{fig:set_d_chart}). Due to the stochastic nature of natural amenity distribution, some runs provide better \enquote{second-best} options for the agents than others.

\unskip

\subsection{Sensitivity Analysis}

While Set D is interesting, it of course may be difficult or even undesirable for policymakers and planners to focus on amenity dispersion and relocation radii in response to displacement: the amenity (such as a harbor) may be inherently undispersable, and relocating the poor away from urban centers certainly may be unjust. Of greater interest are two key parameters---income inequality and density---which decision makers can try to influence through policy. We conduct a sensitivity analysis to test how these parameters influence the state variables---in particular the diversity indicator and the poor access indicator---ceteris paribus to see if they behave as expected from theory cf. \citep{amblard_assessment_2007}. We first hold all the parameters at their baselines values (Set A in Table \ref{tab:parameterizations}), then vary density and income inequality in 20 equal increments one at a time across the parameter space (Table \ref{tab:sensitivity_paramters}). Each simulation is run 50 times, and means and 95\% confidence intervals are calculated for the values at which the model converges.

\begin{table}[h]
	\centering
	\caption{Parameter values for each range of configurations of the simulations in the sensitivity analysis.}
	\label{tab:sensitivity_paramters}
	\small
	\begin{tabular}{lllll}
		\toprule
		& \textbf{Max Dens} & \textbf{Poor Inc} & \textbf{Mid Inc}& \textbf{Rich Inc} \\
		\midrule
		Low dens--Low ineq       & 4      & 40 & 50  & 60  \\
		Low dens--Med ineq    & 4      & 20 & 50  & 100 \\
		Low dens--High ineq      & 4      & 5  & 100 & 200 \\
		Med dens--Low ineq    & 25     & 40 & 50  & 60  \\
		Med dens--Med ineq & 25     & 20 & 50  & 100 \\
		Med dens--High ineq   & 25     & 5  & 100 & 200 \\
		High dens--Low ineq      & 100    & 40 & 50  & 60  \\
		High dens--Med ineq   & 100    & 20 & 50  & 100 \\
		High dens--High ineq     & 100    & 5  & 100 & 200 \\
		\bottomrule
	\end{tabular}
\end{table}

\begin{figure}[h]
	\centering
	\includegraphics[width=1.00\textwidth]{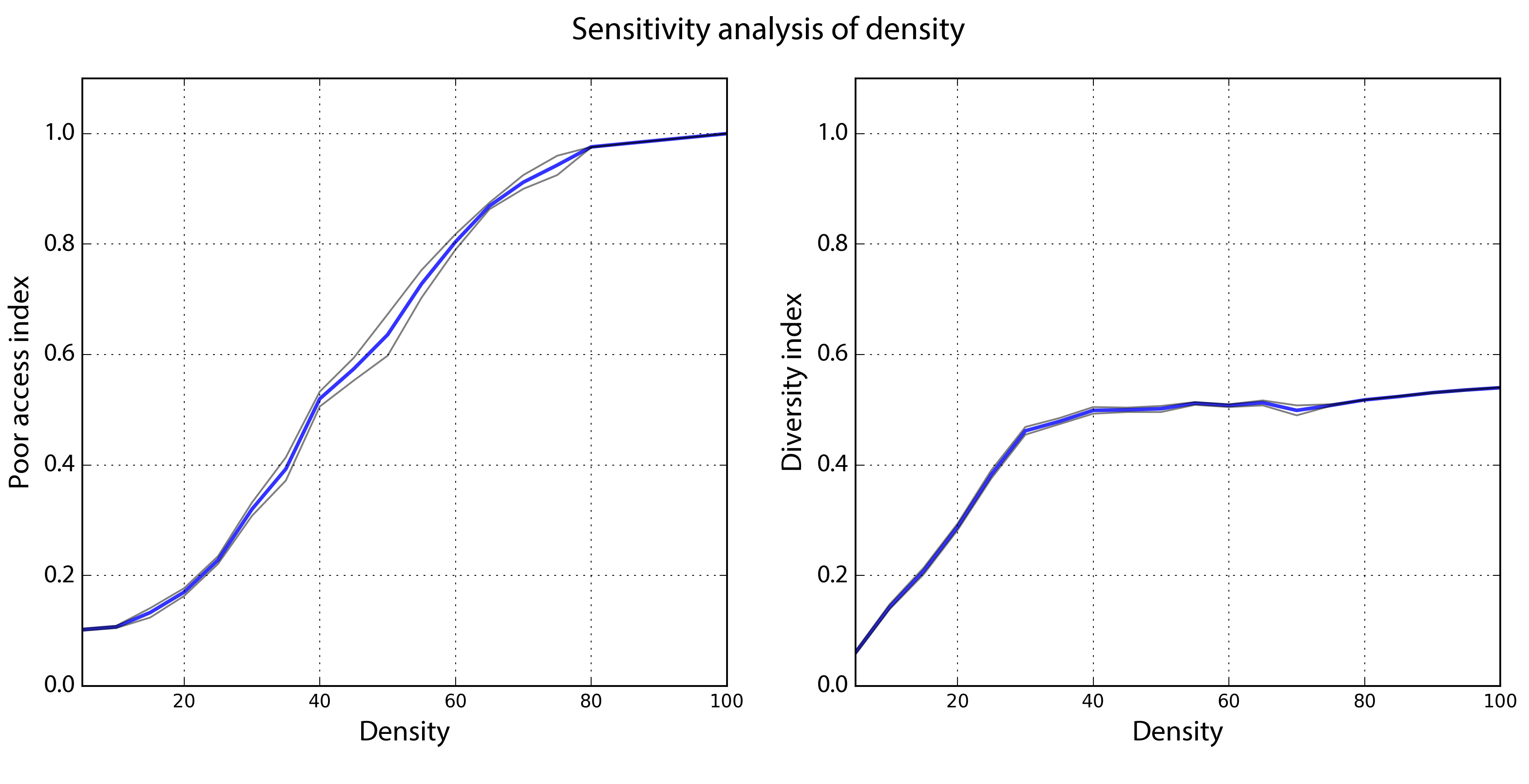}
	\caption{The values for poor access indicator and diversity indicator on which the model converged over 50 runs given different density parameter values, with all other parameters held at their baseline values.  The mean value is blue and the gray lines depict the 95\% confidence interval.}
	\label{fig:sensitivity_density}
\end{figure}

\begin{figure}[h]
	\centering
	\includegraphics[width=0.9\textwidth]{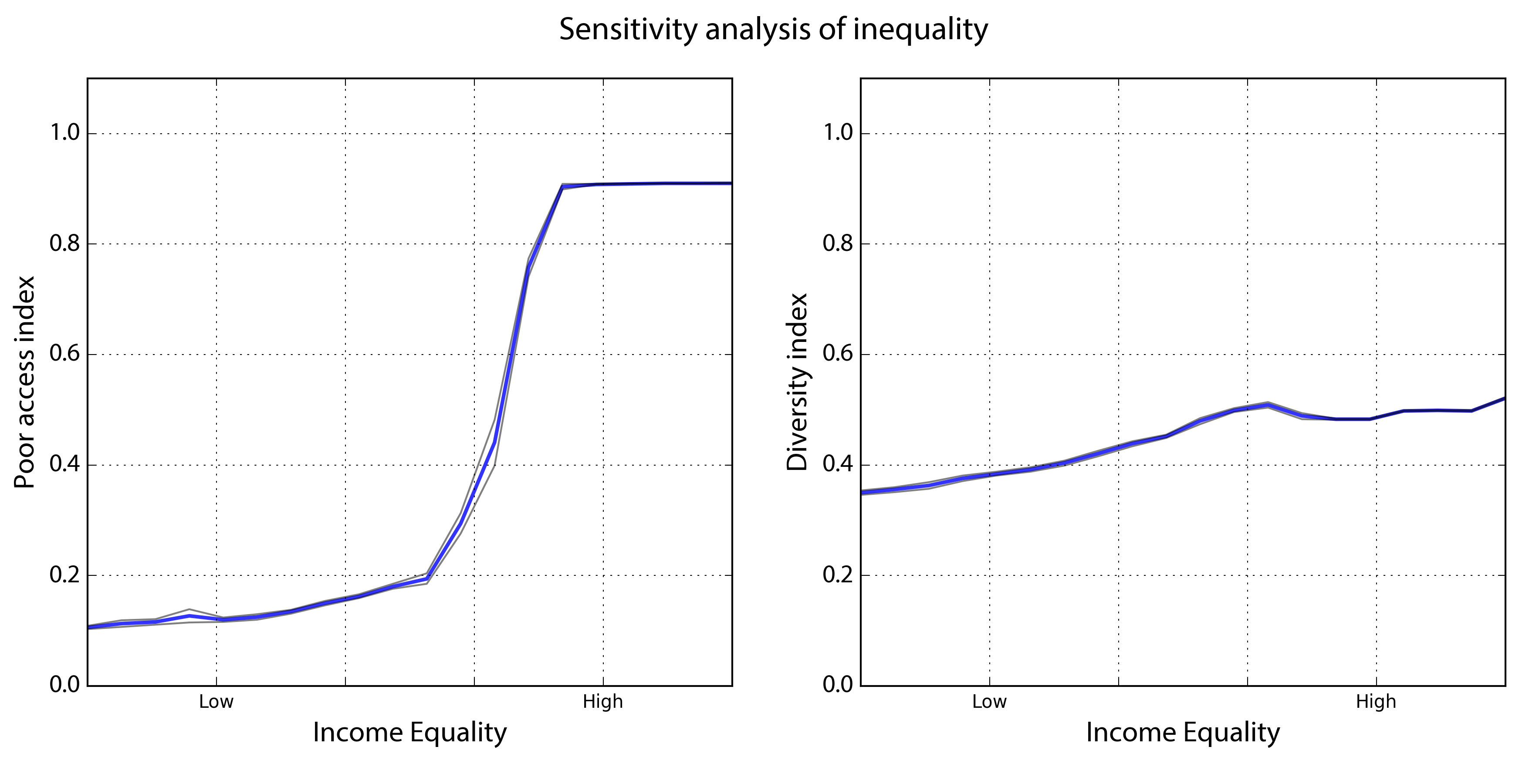}
	\caption{The values for poor access indicator and diversity indicator on which the model converged over 50 runs given different income inequality parameter values, with all other parameters held at their baseline values. The mean value is blue and the gray lines depict the 95\% confidence interval.}
	\label{fig:sensitivity_inequality}
\end{figure}

The poor access indicator increases monotonically in the form of a sigmoid curve as density increases, ceteris paribus. With the other parameters at their baseline values, it increases from a low of 0.11 to a high of 1.00 as density increases (Figure \ref{fig:sensitivity_density}). The diversity indicator also rises with density, but plateaus much earlier than the poor access indicator. It increases from a low of 0.08 to a high of 0.55 as density increases, ceteris paribus. For visual consistency, income inequality is represented here as its inverse: income equality. The poor access indicator again increases monotonically in the form of a sigmoid curve as income equality increases, ceteris paribus. With the other parameters at their baseline values, it increases from a low of 0.11 to a high of 0.91 as income equality increases (Figure \ref{fig:sensitivity_inequality}). The diversity indicator also rises with income equality, albeit less so, increasing from a low of 0.35 to a high of 0.52 as income equality increases, ceteris paribus.

\unskip

Next we segment each parameter into value ranges to explore interactions across the parameter space. Each parameter is set systematically to a low, medium, and high configuration to explore different interacting combinations of density and inequality. We hold the single amenity and relocate radius parameters constant at their baseline values. Earlier we described the selection of parameter values to define low density, medium density, medium inequality, and high inequality. Here we use a density of 100 dwelling units per acre to represent high density, corresponding to mid-rise and high-rise housing. We parameterize low inequality with a poor income of 40, a middle income of 50, and a rich income of 60, which would approximate the 41st, 50th, and 58th percentile household incomes (in thousands of US dollars) in the 2010 US Census.

\begin{figure}[h]
	\centering
	\includegraphics[width=0.6\textwidth]{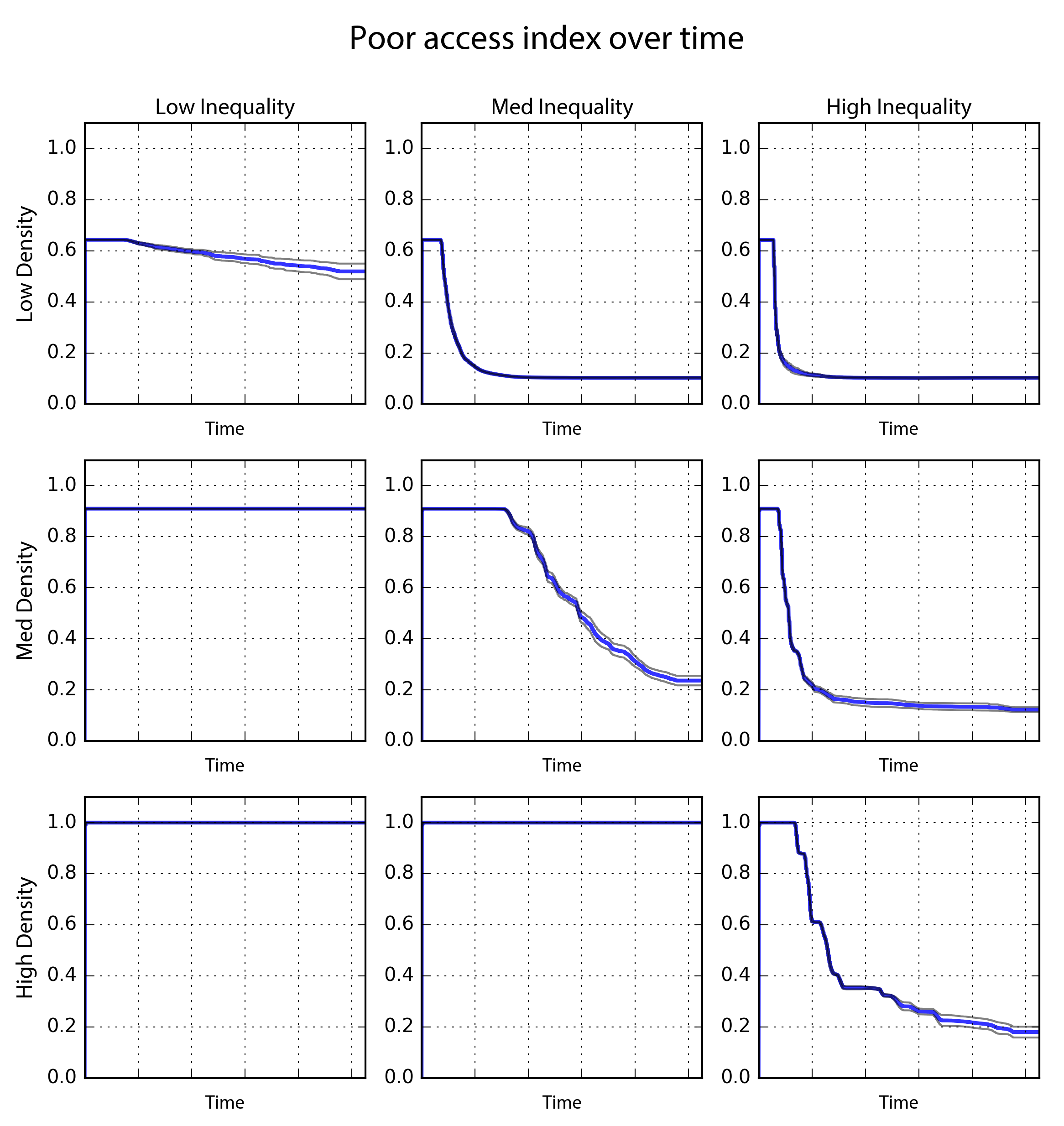}
	\caption{Time series of the poor access indicator values produced by the model when parameterized with varying combinations of low, medium, and high density and inequality values. The mean value across 50 runs is blue and the gray lines depict the 95\% confidence interval.}
	\label{fig:sensitivity_poor_access}
\end{figure}

\begin{figure}[h]
	\centering
	\includegraphics[width=0.6\textwidth]{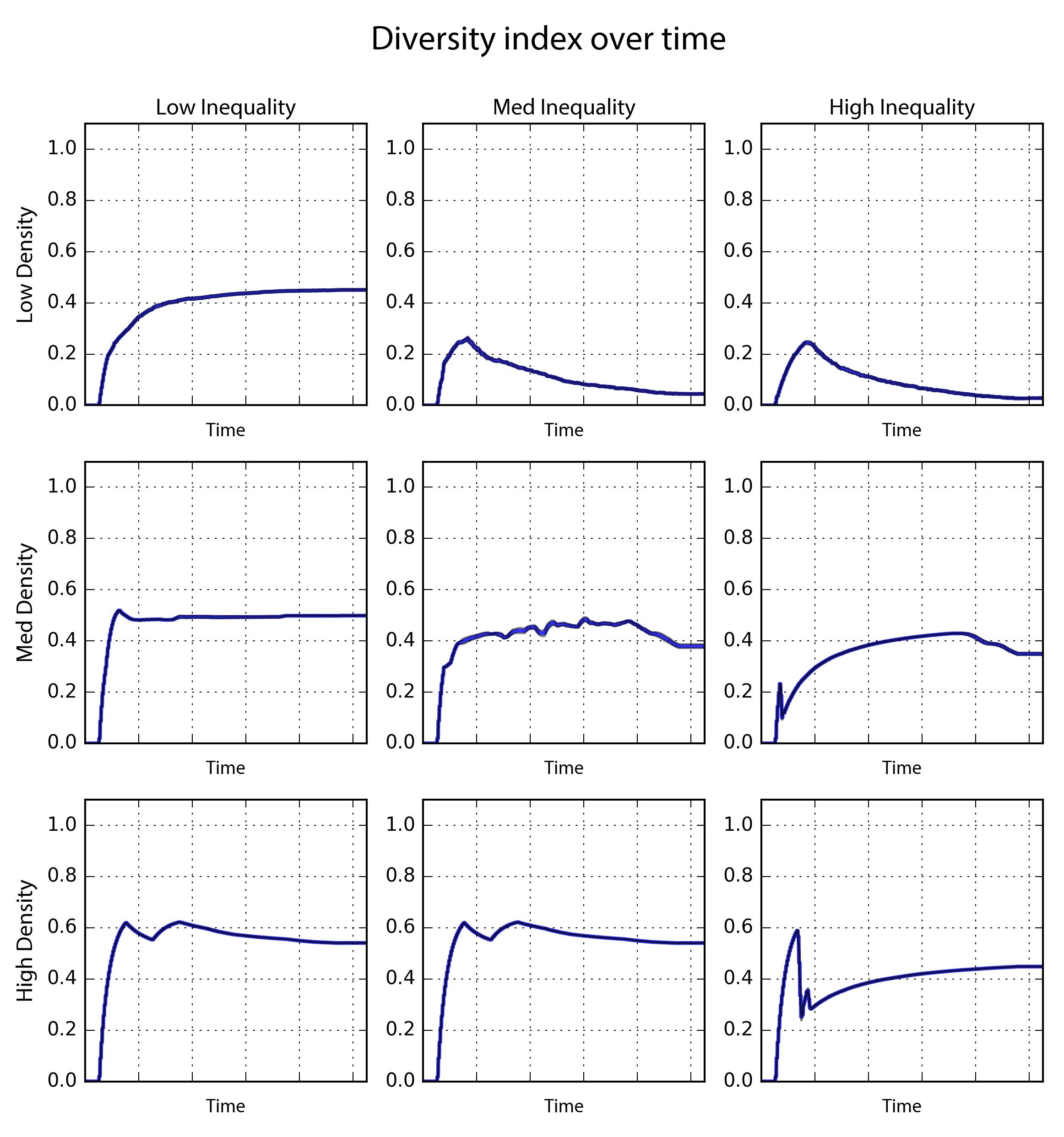}
	\caption{Time series of the diversity indicator values produced by the model when parameterized with varying combinations of low, medium, and high density and inequality values. The mean value across 50 runs is blue and the gray lines depict the 95\% confidence interval.}
	\label{fig:sensitivity_diversity}
\end{figure}

The poor access indicator state variable is sensitive to both parameters (Figures \ref{fig:sensitivity_poor_access} and \ref{fig:sensitivity_diversity}). Poor access is highest when density is high and inequality is low: the high permitted density allows each poor agent to reside on its most preferred cell and the low inequality prevents rich agents from driving up costs and displacing the poor. Poor access is lowest when density is low and inequality is in the medium or high range. In such a scenario, the poor are displaced as costs rise and they are pushed particularly far to the periphery due to the low density restrictions. The only difference in this scenario between medium and high inequality is the speed at which they are displaced. Poor access is low when inequality is high, regardless of density, but fares best with high density. The poor are still displaced, just not as far away.

The diversity indicator state variable is likewise sensitive to both parameters. The model converges on the highest diversity indicator values when parameterized at higher densities and lower inequalities. At high density, it achieves values over 0.40 for all three ranges of inequality, indicating that diversity is reasonably robust to inequality but only in the presence of high density. At high density, poor agents are displaced but only by a cell or two. Thus they are moved off their preferred cell but they remain in the neighborhood. 

At low inequality, the model also produces diversity indicator values over 0.40 for all three ranges of density, indicating that diversity is reasonably robust to density, but only in the presence of low inequality. When inequality is low, density has little impact on diversity as incomes are not stratified enough to stratify settlement patterns by class. Diversity is lowest when density is low and inequality is in the medium or high range. These inequality parameters cause the settlement to stratify by class and the low density pushes agents of other classes outside of one another's Moore neighborhood.

\unskip

\section{Discussion}

This study's toy model demonstrated a simple mechanism by which poor agents could be displaced to the inexpensive periphery of the settlement, dislocated from their initial (and still preferred) settlement location. The rich agents remain clustered around natural amenities they have no interest in, due to path dependence. The system thus evolves to an inefficient state: space could be theoretically redistributed such that poor agents are made better off, without making rich agents worse off. With better planning, the rich agents' urban agglomeration could have been formed a short distance from the natural amenity---in which case the poor would still have concentrated at the periphery of the urban agglomeration, only this time they would also be on their preferred cells.

In the model, spatial displacement occurred even when the displacer had no inherent interest in that point in the landscape or its physical value. Rather, the displacer cared only for the agglomeration of humans that happened to be there. Once that initial agglomeration of agents (who did care about the amenities embedded at that point in the landscape) had been dispersed to the periphery, the displacers remained at the center, forever attracted to their own social construction. This agglomeration stays clustered around the now under-utilized natural amenity because no rich agent has an incentive to move away from the center. 

Thus, a series of individually rational behaviors and preferences aggregate to the societal level in a collectively irrational way. If the group as a whole made a decision, it would be different from the sum of these individual decisions. This illustrates the raison d'être of public planning \citep{friedmann_planning_1987}. \mbox{As \citet{stiftel_planning_2000}} puts it, \enquote{In a market-oriented economy, planning's reason for being is fundamentally tied to this disjunction between individual rationality and collective rationality.} This paper contributes a simplistic mechanism to this literature to illustrate this concisely. The toy model also demonstrated a mechanism by which income equality and density could lessen displacement and improve diversity. When permitted, density was greatest at the center of the agglomeration, which itself was centered on the point of highest natural amenity value. However, the choice to locate at the most preferred cell was restricted in low density zoning schemes that limit people's abilities to economize on space. The poor's access to the natural amenity was highest when income inequality was low and density was high enough to let them settle directly on their preferred cell. Higher inequality and lower densities, by contrast, resulted in the rich displacing the poor to the periphery and away from the resource.

Likewise, diversity was highest when density was high and inequality was low. When inequality is low, the three classes did not stratify themselves into concentric zones by ability to pay, because all had a comparable ability. When density is high, what displacement does occur pushes the poor out only a short distance. This improves poor access and diversity by allowing all three classes to reside next to each other, even if the individual cells are segregated. Density alone did not prevent displacement in this model: only in concert with higher income equality did it yield rich diversity and high accessibility for the poor. This toy model, however, portrays a merely theoretical mechanism and requires validation (through case study research and estimating empirical models) to draw more meaningful conclusions.

Finally, diversity increases as the rich and middle-income agents first move in, but then it drops as they displace the poor and stratify themselves by class into concentric zones. The model's resulting socioeconomic spatial patterns are similar to those seen in wealth-concentrating cities like San Francisco and London as well as cities outside of the developed world, where the rich cluster in metropolitan centers and the poor reside at the periphery.  Interestingly, the poor agents in this model that cluster in mid-sized towns around mid-tier natural amenities do not get displaced. Their urban agglomerations are too small to appeal to the rich who prefer larger, wealthier agglomerations elsewhere. A hierarchy of towns thus emerges from the diverse and recurrent interactions between agents and groups of agents, a finding that echoes researchers studying city hierarchies \citep{pumain_hierarchy_2006,cottineau_growing_2015}.

\section{Conclusions}

This paper presented an ABM of urban settlement, agglomeration, and displacement. It used NetLogo to explore the dynamics of settlement around natural amenities, subsequent settlement around the resulting human agglomeration, displacement of the poor through class succession, and the path dependence of the rich remaining clustered around natural amenities they have no interest in. This demonstrated a simple mechanism by which this type of displacement and path dependence can arise when agents' income levels correlate with amenity-type preferences. The system evolved to an inefficient state: when moving costs are nil, poor agents could be made better off without making rich agents any worse off---if they were to collectively, simultaneously relocate as a group. Public planning could have instead anticipated these effects and made proactive decisions in advance. We found that as density and income equality increase, community diversity increases. Further, density and income equality work synergistically. Higher densities did not improve diversity and the accessibility of the poor unless they were accompanied by greater income equality. Finally, the poor agents that cluster in mid-sized towns around mid-tier natural amenities tend not to get displaced.

In \citet{axtell_agent-based_1994}'s classification system, this is a level 0 model with some level 1 characteristics. It is a simplified conceptual model to caricature a real-world process and illustrate a complex mechanism. However, in terms of validation, it generally conforms qualitatively to real-world macrostructures and spatial distributions cf. \citep{zuk_gentrification_2018,torrens_modeling_2007,zuk_housing_2016}. Thus it offers some potential insights, with a caveat that future research is required to empirically confirm model findings. Increasing income equality might improve the diversity of communities---this is perhaps obvious and well-known in policy circles. However, this model demonstrates strong interaction effects between equality and density. In this model, the mid-sized towns tended to be the most equitable in terms of diversity and access to amenities. Planners might aim for some of these characteristics in larger cities by promoting affordable housing and high-accessibility neighborhoods with a diversity of housing types and amenities.

Future research can extend this model by incorporating wealth-building dynamics and urban dis-economies of scale. Poor agents could harvest natural amenities and increase their wealth and incomes accordingly. Middle-income and rich agents could increase their wealth as a function of proximity to other agents, playing into the logic of urban agglomeration's economic appeal. Additionally, a single agent could have separate variables defining its attraction to natural and urban amenities, rather than the all-or-nothing approach in this model. Currently, the model assumes uniform maximum density across the city, but future work could incorporate zones permitting various maximum densities in different sub-areas. Nevertheless, this model begins to explore these dynamics of the power of path dependence in displacement.

\end{document}